\documentclass[journal]{IEEEtran}

\ifCLASSINFOpdf
\else
   \usepackage[dvips]{graphicx}
\fi
\usepackage{url}

\usepackage{acronym}

\usepackage{cite}
\usepackage[pdftex]{graphicx}
\usepackage{epstopdf}
\graphicspath{{Figures/}}
\usepackage[cmex10]{amsmath}
\usepackage[caption=false]{subfig}
\usepackage{pifont}

\usepackage{stfloats}
\usepackage{url}
\usepackage{amsmath,amsfonts,amssymb,epsfig}
\usepackage{amsthm}
\newcounter{opteq}

 \usepackage[inline]{enumitem}

\usepackage{tikz}
\usetikzlibrary{tikzmark,calc,decorations.pathreplacing,patterns}
\usepackage{tikzsymbols}

\usepackage{pgfplots} 
\pgfplotsset{compat=1.17}
\pgfplotsset{
        colormap={bwrev}{
            rgb255=(255,255,255)
            rgb255=(0,0,0)
        },
        every axis/.append style={
                    label style={font=\small},
                    tick label style={font=\small},  
                    title style={font=\small}, 
                    },
    contour/label node code/.code={%
        \node{$m=\pgfmathprintnumber{#1}$};}
    }
\usepgfplotslibrary{fillbetween}
\usetikzlibrary{pgfplots.groupplots}
\def\tikz@auto@anchor{%
    \pgfmathtruncatemacro\angle{atan2(\pgf@x,\pgf@y)-90}
    \edef\tikz@anchor{\angle}%
}
\usepackage{xfp}

\usepackage{xcolor}

\usepackage{hyperref}
\usepackage[capitalize]{cleveref}

\newtheorem{definition}{Definition}
\newtheorem{lemma}{Lemma}

\newtheorem{thm}{Theorem}
\newtheorem{corollary}{Corollary}
\newtheorem{remark}{Remark}

\Crefname{thm}{Theorem}{Theorems}
\crefname{thm}{Theorem}{Theorems}

\usepackage{algorithm}
\usepackage{algorithmicx}
\usepackage[noend]{algpseudocode}

\usepackage{mathtools}
\usepackage{bbm}
\usepackage{arydshln}
\usepackage{mathrsfs}
\usepackage{bm}
\usepackage{lipsum}

\usepackage{threeparttable}
\usepackage{multirow}
\usepackage{multicol}
\usepackage[makeroom]{cancel}
\usepackage{adjustbox}

\newcommand{\txt}[1]{\text{\normalfont #1}}

\DeclareMathOperator{\HT}{H}
\DeclareMathOperator{\F}{F}
\DeclareMathOperator{\rank}{rank}
\DeclareMathOperator{\krank}{k-rank}
\DeclareMathOperator{\diag}{diag}

\DeclareMathOperator{\kron}{\otimes}
\DeclareMathOperator{\krao}{\odot}

\DeclareMathOperator{\tx}{t}
\DeclareMathOperator{\rx}{r}

\DeclareMathOperator{\aiii}{\labelcref{i:array_bad_gma}}
\DeclareMathOperator{\aiv}{\labelcref{i:array_good_2}}
\DeclareMathOperator{\athree}{\labelcref{i:array_bad_3}}
\DeclareMathOperator{\afour}{\labelcref{i:array_good_4}}
\DeclareMathOperator{\nrs}{\mathbb{S}_{\txt{NRS}}}

\definecolor{ForestGreen}{HTML}{4DCB42}

\acrodef{ula}[ULA]{uniform linear array}
\acrodef{mimo}[MIMO]{multiple-input multiple-output}
\acrodef{nrs}[NRS]{nonredundant singleton subset}
\acrodef{snr}[SNR]{signal-to-noise ratio}
\acrodef{isac}[ISAC]{integrated sensing and communication}
\acrodef{dof}[DoFs]{degrees-of-freedom}
\acrodef{doa}[DoA]{direction-of-arrival}
\acrodef{fft}[FFT]{fast Fourier transform}
\acrodef{gma}[GMA]{Generalized MIMO array}
\acrodef{mmwave}[mmWave]{millimeter wave}
\acrodef{mse}[MSE]{mean squared error}
\acrodef{nextg}[Next-G]{next-generation}
\acrodef{kpi}[KPI]{key performance indicator}
\acrodef{tx}[Tx]{transmit}
\acrodef{rx}[Rx]{receive}
\acrodef{rsc}[RSC]{redundancy subspace condition}
\acrodef{rf}[RF]{radio-frequency}
\acrodef{rmse}[RMSE]{root-mean-squared error}
\acrodef{rw}[RW]{redundancy-waveform}
\acrodef{wr}[WR]{waveform rank}

\crefname{enumi}{Array}{Arrays}

\begin{document}

\title{On the Impossibility of Lossless Waveform Rank Reduction for Certain Redundant Arrays}

\author{Robin Rajam\"{a}ki, \IEEEmembership{Member, IEEE}, Piya Pal, \IEEEmembership{Senior Member, IEEE}
\thanks{R. Rajam\"{a}ki is with the Signal Processing Research Centre, Tampere University, Finland (e-mail: robin.rajamaki@tuni.fi). He was previously with the Department of Information and Communications Engineering, Aalto University, Finland.}
\thanks{P. Pal is with the Electrical and Computer Engineering department, University of California San Diego, USA (e-mail: pipal@ucsd.edu).}}

\maketitle

\begin{abstract}
Efficient use of spatio-temporal resources, including sensor arrays and transmit waveforms, is a key challenge in modern \acs{mimo} active sensing systems. This paper studies the impact of array redundancy and \acf{wr} on active sensing performance. Specifically, we show that parameter identifiability at \emph{reduced} \ac{wr} critically depends on subspace properties of the so-called array redundancy pattern. 
We show that array geometries with identical sum co-arrays can exhibit markedly different identifiability properties at low \ac{wr}. We derive a novel necessary condition for maximizing identifiability at reduced \ac{wr}, which reveals that the unfavorable redundancy patterns of certain redundant arrays fundamentally limits their performance. The results yield new insights into resource-efficient sensing systems, motivating redundancy-aware array and waveform design.
\end{abstract}


\IEEEpeerreviewmaketitle


\section{Introduction}\label{sec:intro}
\acresetall


\Ac{mimo} active sensing systems, such as \ac{mimo} radars, leverage the ability of transmitters to launch independent \emph{waveforms} to improve sensing performance, including target identifiability and resolution. Such \ac{mimo} 
waveforms can be represented by a $T\times N_{\tx}$ 
matrix $\bm{S}
$, where $N_{\tx}$ denotes the number of \ac{tx} sensors and $T$ is the waveform length (in samples). It is well-known that \emph{full-rank waveforms} ($\rank(\bm{S})=N_{\tx}$), 
such as orthogonal waveforms, 
reap the full diversity benefits provided by the $N_{\tx}$ transmitters \cite{li2007onparameter}. 
However, \emph{reduced-rank} waveforms ($\rank(\bm{S})<N_{\tx}$) also have several advantages, such as improved transmit beamforming gain and resulting parameter estimation performance \cite{li2008range,hassanien2011transmit}, and a need for fewer costly \ac{rf} chains. Low-dimensional \ac{tx} signals also naturally arise in 
modern \ac{mimo} communications systems employing an excess of base station antennas compared to the number of user devices \cite{ngo2024ultradense}. 
With the emergence of \ac{isac} \cite{paul2017survey,liu2023integrated,ahmadipour2024aninformation,koivunen20204multicarrier}, reduced \ac{wr} is thereby increasingly relevant also in sensing. 
This begs the questions: 
\emph{Can \ac{wr} be reduced 
without giving up  
the 
performance 
of 
full \ac{wr}?} 
\emph{What enables or prevents doing this?}

To answer these questions, we will shed new light on the \emph{sum co-array} \cite{hoctor1990theunifying}---a ubiquitous virtual array model in active sensing, known to govern sensing performance at full \ac{wr}. Specifically, we will show that when operating at reduced \ac{wr}, \emph{not all \ac{mimo} arrays with the same co-array are created equal}. Indeed, there exists array geometries for which reducing \ac{wr} leads to an inevitable loss of performance, such as fewer identifiable targets. Our focus is on \emph{redundant} array geometries, which have repeated virtual sensors---in contrast to nonredundant arrays typically considered in \ac{mimo} radar. Redundant arrays, such as \acp{ula}, are key in a multitude of applications prioritizing, e.g., beamforming capability \cite{hassanien2011transmit,rajamaki2020hybrid,liu2023integrated} and robustness to sensor failure
\cite{liu2019robustnessi}. 
An important and insofar largely neglected attribute of the array geometry is the so-called \emph{redundancy pattern} (defined in \cref{sec:sumcoarray}), which provides a mapping between the \emph{physical} \ac{tx} and \ac{rx} sensors and the \emph{virtual} sensors of the co-array. 
As we will see, the redundancy pattern plays a key role in reducing \ac{wr} without sacrificing performance. This fact has several surprising consequences, including that 
array geometries 
with identical co-arrays and physical sensor counts can have different identifiability properties at reduced \ac{wr}.

\emph{Contributions:}
This paper establishes that certain redundant arrays cannot reduce \acl{wr} without sacrificing identifiability. Specifically, we show how the ability to operate at low \ac{wr} depends on the subspace properties of the redundancy pattern, and that its combinatorial and binary structure renders characterizing these properties highly nontrivial. We then derive a novel necessary condition for maximizing identifiability at reduced \ac{wr}. This condition pinpoints a general structure in the co-array that hampers identifiability at low \ac{wr}. 
Finally, we validate our theoretical findings via illustrative 
simulations.

\emph{Related work:} 
Sparse arrays \cite{wang2017coarrays,sarangi2023superresolution,amin2024sparsearrays} and waveform design \cite{bell1993information,levanon2004radar,he2012waveform} 
have attracted significant research interest, most recently in the context of \ac{isac} \cite{wang2019dual,sankar2024sparse,vanderwerf2024receiver,rajamaki2024sparse}. 
For example, \cite{sankar2024sparse} considered a precoder design problem with rank and identifiability constraints. 
However, leveraging array redundancy to reduce precoder rank with identifiability guarantees was not explored. We fill this gap by providing new insights into low-rank waveform design and its dependence on the redundancy pattern of the array geometry. Identifiability in multisensor systems has also been widely studied \cite{bresler1986onthe,wax1989onunique,nehorai1991direction,abramovic1999resolving,yang2018sparsemethods}. 
For instance, \cite{chowdhury2026direction} recently revisited the identifiability of (oversampled) \acp{ula}---the canonical redundant array---in passive sensing.
Works on active sensing also exist \cite{li2007onparameter,wang2011onparameter,shulkind2016direction,hu2019parameter,shi2022onparameter}, but typically assume full-rank waveforms or do not provide rigorous identifiability conditions. Our results address this limitation by 
revealing the impact of low-rank transmit waveforms on identifiability when employing redundant array geometries, and showing when lossless \ac{wr} reduction is impossible. Our preliminary work \cite{rajamaki2023importance} showcased a subset of the ideas developed in this paper by providing an isolated example of arrays failing at reduced \ac{wr}. This manuscript significantly generalizes \cite{rajamaki2023importance} and our earlier preprint \cite{rajamaki2023arrayinformed} by deriving a novel necessary condition for redundant arrays to operate at reduced \ac{wr}. 
This so-called nonredundant singleton subset (NRS) condition was not identified in \cite{rajamaki2023importance,rajamaki2023arrayinformed}.  
The manuscript also contains multiple numerical examples illustrating the theory, which were not included in \cite{rajamaki2023importance,rajamaki2023arrayinformed}.

\section{Signal model and problem formulation}
We consider a narrowband \ac{mimo} active sensing system estimating the unknown \acp{doa} $\bm{\theta}\in[-\frac{\pi}{2},\frac{\pi}{2})^K$ of $K$ targets in the same delay-Doppler cell. Given $N_{\rx}$ \ac{rx} sensors and $T$ temporal samples, the 
received backscatter signal 
can be modeled as $N_{\rx}T\!\times\!1$ 
vector \cite{bekkerman2006target,friedlander2012onsignal}
\begin{align}
\bm{y}
\!=\!
\bm{B}(\bm{\theta})\bm{x}\!+\!\bm{n},
\  \text{where}\  
\bm{B}(\bm{\theta})\!\triangleq\!
(
\bm{S}
\bm{A}_{\mathbb{D}_{\tx}}(\bm{\theta}))\krao
\bm{A}_{\mathbb{D}_{\rx}}
(\bm{\theta}).
\label{eq:measurementmodel}
\end{align}
Here, $\bm{S}\in\mathbb{C}^{T\times N_{\tx}}$ is a known 
\emph{spatio-temporal \ac{tx} waveform matrix}, $N_{\tx}$ is the number of \ac{tx} sensors, $ \bm{x} \in\mathbb{C}^K $ are the 
target scattering coefficients, $ \bm{n} \in\mathbb{C}^{N_{\rx}} $ a noise vector, and $\krao$ denotes the Khatri-Rao (columnwise Kronecker) product. 
Moreover, 
$
\bm{A}_{\mathbb{D}_{\tx}}(\bm{\theta})\!\in\!\mathbb{C}^{N_{\tx}\times K}$ and $
\bm{A}_{\mathbb{D}_{\rx}}(\bm{\theta})\!\in\!\mathbb{C}^{N_{\rx}\times K}$ are the \ac{tx} and \ac{rx} array manifolds, respectively. 
Under standard ideal sensor assumptions, the $(n,k)$the element of manifold matrix $\bm{A}_{\mathbb{D}}(\bm{\theta})$ of array $\mathbb{D}\!=\!\{d[n]\}_{n=1}^N$ (sensor positions are assumed normalized to units of half a wavelength) can be written as
\begin{align}
 [\bm{A}_{\mathbb{D}}(\bm{\theta})]_{n,k}=\exp(j\pi d[n] \sin \theta_k).   \label{eq:A}
\end{align}
\cref{eq:A} enables decomposing 
$\bm{B}(\bm{\theta})$ in \eqref{eq:measurementmodel}, henceforth referred to as the 
\emph{spatio-temporal sensing matrix}, as follows \cite{rajamaki2023arrayinformed}: 
\begin{align}
   \bm{B}(\bm{\theta}) 
   =\bm{W}
   \bm{A}_{\mathbb{D}_\Sigma}(\bm{\theta}), \text{ where } \bm{W}\triangleq (\bm{S} \kron\bm{I}_{N_{\rx}})\bm{\Upsilon}.\label{eq:B}
\end{align}
Here, 
$
\bm{A}_{\mathbb{D}_\Sigma}(\bm{\theta})\in\mathbb{C}^{N_\Sigma \times K}$ is the manifold of  
the \emph{sum co-array} $\mathbb{D}_\Sigma$, 
defined as the sum set of \ac{tx} and \ac{rx} sensor position pairs:
\begin{align}
    \mathbb{D}_\Sigma  \triangleq \mathbb{D}_{\tx}+\mathbb{D}_{\rx}=
    \{d_{\tx} +d_{\rx}\ |\ d_{\tx}\in\mathbb{D}_{\tx}; d_{\rx}\in\mathbb{D}_{\rx}\}.
    \label{eq:sca}
\end{align}
Matrix $\bm{W}$ in \eqref{eq:B} is a function of the so-called \emph{redundancy pattern} $\bm{\Upsilon}\in\{0,1\}^{N_{\tx}N_{\rx}\times N_\Sigma}$ \cite{rajamaki2020hybrid}, which is a binary matrix mapping the unique elements of the sum co-array to the physical \ac{tx}-\ac{rx} sensor pairs. 
\cref{sec:sumcoarray} will return to this key quantity 
in detail. In the remainder of this paper, we will assume that the sum co-array is \emph{contiguous}, i.e., $\mathbb{D}_\Sigma$ consists of $N_\Sigma$ consecutive integers. 

{\em Problem formulation:} We are interested in the following question: {\em ``Can the rank of $\bm{B}(\bm{\theta})$ be maintained at $N_\Sigma$ (its maximum possible value) for all choices of $\bm{\theta}$, with a waveform matrix $\bm{S}$ of rank less than $N_{\tx}$?"} 
Indeed, if $\rank(\bm{B}(\bm{\theta}))=N_\Sigma$ for all $\bm{\theta}$ ($\theta_k\!\neq\!\theta_i, \forall k\!\neq\!i$), 
then it can be shown that up to $K\!=\!N_\Sigma/2$ \acp{doa} can be uniquely identified \cite{yang2018sparsemethods}.\footnote{That is, no $\bm{\theta}'\neq \bm{\theta}$ can give rise to the same (noiseless) measurement as $\bm{\theta}$, such that $\bm{B}(\bm{\theta})\bm{x}\neq\bm{B}(\bm{\theta}')\bm{x}'$ for all distinct $\bm{\theta},\bm{\theta}'$ and nonzero $\bm{x},\bm{x}'$.} 
Our goal is to investigate if this can be ensured {\em even with a waveform matrix with reduced WR}. As we will establish in the next section, the redundancy pattern $\bm{\Upsilon}$ of the Tx-Rx array geometries plays an important and subtle role in determining if this is possible.


\section{Role of redundancy subspace in waveform rank reduction
}\label{sec:sumcoarray}

The redundancy pattern matrix $\bm{\Upsilon}$ 
encodes which of the $N_{\tx}N_{\rx}$ physical Tx-Rx sensor pairs contribute to which of the  $N_\Sigma\!\leq\!N_{\tx}N_{\rx}
$ unique virtual sensors per 
the following definition.
\begin{definition}[Redundancy pattern]\label{def:Upsilon}
The ($i,\ell$)th entry of redundancy pattern matrix $ \bm{\Upsilon}\in\{0,1\}^{N_{\tx}N_{\rx}\times N_\Sigma} $ is defined as
\begin{align}
\Upsilon_{i,\ell}\triangleq
    \begin{cases}
       1,& \txt{if } d_{\tx}[n]+d_{\rx}[m]=d_\Sigma[\ell]\\
       0,&\txt{otherwise},
    \end{cases}\label{eq:Upsilon}
\end{align}
where $i=i[{n,m}]\triangleq m+(n-1)N_{\rx}$. 
That is, $\Upsilon_{i,\ell}$ is unity if Tx-Rx pair $i$ contributes to virtual sensor $\ell$ (zero otherwise).
\end{definition}
The columns of $\bm{\Upsilon}$ encode the redundancies of the sum co-array elements in $\mathbb{D}_\Sigma$ (which is a set with nonrepeated elements). 
For a nonredundant array ($N_{\tx}N_{\rx}=N_\Sigma)$, $\bm{\Upsilon}$ is a (square) permutation matrix, since, by definition, each column (in addition to each row) contains only one nonzero entry. 
In this case, full \ac{wr} is necessary to maximize identifiability. Indeed, if $\rank(\bm{S})<N_{\tx} $ and $N_\Sigma=N_{\tx}N_{\rx}$, then $\rank(\bm{B}(\bm{\theta}))\leq
\rank(\bm{W})
\leq\rank(\bm{S}\kron\bm{I}_{N_{\rx}})=\rank(\bm{S})\rank(\bm{I}_{N_{\rx}})
<N_{\tx}N_{\rx}=N_\Sigma$. This establishes 
that for nonredundant arrays, it is impossible to reduce \ac{wr} while maintaining $\rank(\bm{B}(\bm{\theta}))=N_\Sigma$. Therefore, a \emph{necessary} condition for \ac{wr} reduction is that the sum co-array be redundant ($N_\Sigma<N_{\tx}N_{\rx}$). But is this sufficient? 
\cref{sec:nonexistence} will show that, interestingly, the answer is \emph{no}. To understand why, we first establish that the \emph{subspace} spanned by the columns of $\bm{\Upsilon}$ controls if a redundant array can achieve maximum identifiability at reduced \ac{wr}.
\begin{lemma}[\Acl{rsc}]\label{thm:gen_restr_iff}
Given 
a contiguous sum co-array with redundancy pattern $\bm{\Upsilon}$, 
one can ensure $\rank(\bm{B}(\bm{\theta}))=N_\Sigma$ for all $\bm{\theta}$ using a rank-deficient waveform matrix $\bm{S}$ ($\rank(\bm{S})<N_{\tx}$) 
if and only if $\rank((\bm{S}\kron\bm{I})\bm{\Upsilon})=N_\Sigma$, i.e., if and only if
  \begin{align}
        \dim\big( \mathcal{N}(\bm{S}\kron\bm{I})\cap \mathcal{R}(\bm{\Upsilon})\big)= 
        0.\label{eq:rank_nullity}
    \end{align}
    We refer to \eqref{eq:rank_nullity} as the \underline{``\ac{rsc}''}.
\end{lemma}
\begin{proof}
We introduce the notion of Kruskal rank \cite{kruskal1977threeway}, which enables writing $\rank(\bm{B}(\bm{\theta}))=N_\Sigma$ \emph{for all} $\bm{\theta}$ equivalently (and compactly) as $\krank (\mathbb{B})=N_\Sigma$.
Here, 
$\mathbb{B}\!\subseteq\!\mathbb{C}^{N_{\rx}T}$ 
denotes 
the set of possible columns of $\bm{B}(\cdot)$ in \eqref{eq:B}:
\begin{align}
    \mathbb{B}\triangleq\{\bm{B}(\phi)\ |\  \phi\!\in\![-\tfrac{\pi}{2},\tfrac{\pi}{2})\}.\label{eq:B_set}
\end{align}
The Kruskal rank of $\mathbb{B}$, denoted $\krank(\mathbb{B})$, is the largest integer $\ell$, such that \emph{any} subset of $\ell$ \emph{unique} elements of $\mathbb{B}$ are linearly independent. Any $K$ angles $\bm{\theta}\in[-\frac{\pi}{2},\frac{\pi}{2})^K$ are identifiable if and only if $K\!\leq\!\tfrac{1}{2}\krank(\mathbb{B})$ \cite[Theorem~11.1]{yang2018sparsemethods}. 
Showing that $\rank(\bm{B}(\bm{\theta}))=N_\Sigma$ for all $\bm{\theta}$ ($\theta_k\neq\theta_i, \forall k\neq i$) is therefore equivalent to establishing that $\krank(\mathbb{B})=N_\Sigma$.

Before specializing our result to the case of contiguous sum co-arrays, we prove the following generalization of \cref{thm:gen_restr_iff}:
\begin{align}
    \krank(\mathbb{B})=N_\Sigma\iff
    \begin{cases}
    \rank((\bm{S}\kron\bm{I})\bm{\Upsilon})=N_\Sigma\\
    \krank(\mathbb{A}_\Sigma)=N_\Sigma.
    \end{cases}\label{eq:rwm_general}
\end{align}
Here, $\mathbb{A}_\Sigma\triangleq \{\bm{A}_{\mathbb{D}_\Sigma}(\phi)\ |\  \phi\!\in\![-\tfrac{\pi}{2},\tfrac{\pi}{2})\}\subseteq\mathbb{C}^{N_\Sigma}$ denotes the set of possible columns of co-array manifold matrix $\bm{A}_{\mathbb{D}_\Sigma}(\cdot)$ in \eqref{eq:B}. 
By \eqref{eq:B}, $\bm{B}(\bm{\theta})\!=\!\bm{W}\bm{A}_{\mathbb{D}_\Sigma}(\bm{\theta})$, where $\bm{W}\!=\!(\bm{S}\kron\bm{I})\bm{\Upsilon}$. 
By definition, the Kruskal rank is invariant to linear transforms that preserve the null space. That is, 
if $\rank(\bm{W})\!=\!N_\Sigma$ and $\krank(\mathbb{A}_\Sigma)\!=\!N_\Sigma$, then $\krank(\mathbb{B})\!=\!\krank(\mathbb{A}_\Sigma)\!=\!N_\Sigma$.

Conversely, suppose $\krank(\mathbb{B})\!=\!N_\Sigma$. Then $N_\Sigma\!=\!\krank(\mathbb{B})\!\leq\!\rank(\bm{B}(\bm{\theta}))\!\leq\!\rank(\bm{W})\!\leq\!N_\Sigma$. Hence, all inequalities should hold with equality, i.e., $\rank(\bm{W})\!=\!N_\Sigma$. 
Since $\bm{W}$ preserves the null space of $\bm{B}(\bm{\theta})$, it must hold that 
$\krank(\mathbb{B})\!=\!\krank(\mathbb{A}_\Sigma)\!=\!N_\Sigma$, proving \eqref{eq:rwm_general}.

Finally, when the co-array is contiguous\footnote{In the noncontiguous case, the \ac{rsc} \eqref{eq:rank_nullity} remains necessary, but not sufficient. Identifiability conditions are more challenging to derive for the noncontiguous case (cf. \cite{chen2021rank}), which is hence left for future work.
}, it is well known that $\krank(\mathbb{A}_\Sigma)=N_\Sigma$, since $\bm{A}_{\mathbb{D}_\Sigma}(\bm{\theta})$ is a Vandermonde matrix with distinct columns \cite[p.~345]{stoica2005spectral}. Hence, if $\mathbb{D}_\Sigma$ is contiguous, then $\krank(\mathbb{B})=N_\Sigma$ if and only if $\rank((\bm{S}\kron\bm{I})\bm{\Upsilon})=N_\Sigma$. \cref{eq:rank_nullity} then follows from the rank-nullity theorem:
\begin{align*}
    \rank((\bm{S}\kron\bm{I})\bm{\Upsilon})
    = \rank(\bm{\Upsilon})-\dim\big( \mathcal{N}(\bm{S}\kron\bm{I})\cap\mathcal{R}(\bm{\Upsilon})\big),
\end{align*}
since $\rank((\bm{S}\kron\bm{I})\bm{\Upsilon})\!=\!N_\Sigma\!=\!\rank(\bm{\Upsilon})$ by definition.
\end{proof}
 
\begin{remark}[Challenges in satisfying \ac{rsc}]\label{thm:nontriviality} 
\cref{thm:gen_restr_iff} implies that a redundant sum co-array is not enough for \ac{wr} reduction---rather, the exact redundancy pattern $\bm{\Upsilon}$ matters. In other words, $\bm{\Upsilon}$ has to permit 
the existence of reduced rank $\bm{S}$ satisfying \ac{rsc}. 
Determining this is highly nontrivial, 
mainly due to the specialized combinatorial and binary structure of $\bm{\Upsilon}$, which limits possible choices of $\bm{\Upsilon}$ and thereby its column space. Similarly, Kronecker structure $\bm{S}\!\kron\!\bm{I}$ makes it nontrivial to 
characterize 
rank-deficient $\bm{S}$ satisfying \eqref{eq:rank_nullity} given $\bm{\Upsilon}$.
\end{remark} 
When both $\bm{S}$ and $\bm{\Upsilon}$ are given, the RSC simply reduces to checking the rank of a known matrix. \cref{thm:nontriviality} 
is most strikingly illustrated by the fact that for certain redundancy patterns 
$\bm{\Upsilon}$,  
\emph{no} rank-deficient waveform matrix $\bm{S}$ exists such that \eqref{eq:rank_nullity} holds, as the next section shows. 

\subsection{Necessary condition for \acl{rsc}} \label{sec:nonexistence}
The problem of which redundancy patterns $\bm{\Upsilon}$ satisfy \eqref{eq:rank_nullity} when $\bm{S}$ is rank deficient is currently wide open. As a step towards bridging this gap, we show the existence of a certain set of columns in $\bm{\Upsilon}$ 
violates \eqref{eq:rank_nullity} for \emph{any} rank-deficient $\bm{S}$. 
\begin{definition}[\Acl{nrs}]\label{def:nrs}
A subset $\nrs$ of $\mathbb{D}_{\Sigma}$  is called a \acf{nrs} if it satisfies the following properties: 
\begin{enumerate}
    \item $\vert \nrs\vert =N_{\tx}$.\label{i:nrs_cardinality}
    \item The redundancy of each element of $\nrs$ is $1$.\label{i:nrs_redundancy}
    \item Each element of $\nrs$ can be expressed as the sum of the {\bf same Rx sensor} with one of the $N_{\tx}$ Tx sensors. \label{i:nrs_rx}
\end{enumerate}
That is, an \ac{nrs} is a set of virtual sensors formed by pairing all Tx sensors with one Rx sensor, where none of the resulting virtual sensors overlap with those of other Tx–Rx pairs.
\end{definition}
Property \labelcref{i:nrs_redundancy} of \cref{def:nrs} simply means that the columns of $\bm{\Upsilon}$ corresponding to co-array elements $\nrs$ have a single nonzero entry each. This can be expressed using the multiplicity $\upsilon(d_\Sigma)$ of co-array element $d_\Sigma$ as $\upsilon(s)=1,\forall s\in\nrs$, where $\upsilon(d_\Sigma) \triangleq \sum_{d_{\tx}\in\mathbb{D}_{\tx}} \sum_{d_{\rx}\in\mathbb{D}_{\rx}} \mathbbm{1}(d_{\tx}+d_{\rx}=d_\Sigma)$. Moreover, by Property~\labelcref{i:nrs_rx}, each element $s\in\nrs$ can be written as $s=d_{\tx}+d_{\rx}'$ for some \ac{rx} sensor $d_{\rx}'\in\mathbb{D}_{\rx}$, \emph{fixed} for all $s\in\nrs$, and \ac{tx} sensor $d_{\tx}\in\mathbb{D}_{\tx}$ depending on $s$.

The following theorem proves that for RSC to be satisfied, it is necessary that $\mathbb{D}_{\Sigma}$ does not contain an NRS.
\begin{thm}
\label{thm:unfavorable_set}
    If sum co-array $\mathbb{D}_\Sigma$ contains an \ac{nrs}, 
    then no choice of rank-deficient $\bm{S}$ can satisfy \ac{rsc} \eqref{eq:rank_nullity}. In other words, \ac{rsc} holds only if $\mathbb{D}_{\Sigma}$ does not contain an \ac{nrs}.
\end{thm}
\begin{proof}
We prove that the existence of an \ac{nrs} implies that redundancy subspace condition \eqref{eq:rank_nullity} is violated for all rank-deficient $\bm{S}$. 
For convenience, let $\mathbb{I}_\ell$ denote the set of \ac{tx}-\ac{rx} sensor indices 
contributing to the $\ell$th virtual sensor:
\begin{align}
    \mathbb{I}_\ell
    \triangleq \{ (n,m) \text{ such that } d_{\tx}[n]+d_{\rx}[m]=d_{\Sigma}[\ell] \}. \label{eq:set_redundancy}
\end{align}
Hence, if $(n,m)\in\mathbb{I}_\ell$, then the $\ell$th column of redundancy pattern matrix $\bm{\Upsilon}$ has a nonzero entry on row $i=m+(n-1)N_{\rx}$, by \cref{def:Upsilon}. This indexing allows equivalently expressing 
the $\ell$th column of $\bm{\Upsilon}$ using a sum of Kronecker product:
    \begin{align}
         [\bm{\Upsilon}]_{:,\ell}\!=\! 
        \sum_{(n,m)\in\mathbb{I}_\ell}
         \bm{e}_{n}^{N_{\tx}} \kron \bm{e}_{m}^{N_{\rx}},
         \label{eq:upsilon_vec}
    \end{align}
    where 
    $\bm{e}_p^{P}\in\{0,1\}^P$ is the standard unit vector (of length $P$) whose $p$th entry is unity. 
    Furthermore, let $\mathbb{J}=\nrs +1$ denote the virtual sensor indices of \ac{nrs} elements $\nrs$. 
    This follows from the contiguity of the sum co-array, as the virtual sensors can be indexed as $d_\Sigma[\ell]=\ell-1, \ell = 1,2,\ldots,N_\Sigma$ without loss of generality. 
    By Property~\labelcref{i:nrs_cardinality} of \cref{def:nrs}, we may write $\mathbb{J}\!=\!\{i_n\}_{n=1}^{N_{\tx}}$. By Property~\labelcref{i:nrs_redundancy}, $\mathbb{I}_i$ is a singleton when $i\!\in\!\mathbb{J}$, i.e., $|\mathbb{I}_i|\!=\!1,\forall i\!\in\!\mathbb{J}$. Hence, column $i_n\!\in\!\mathbb{J}$ 
    of $\bm{\Upsilon}$ satisfies
\begin{align*}
    [\bm{\Upsilon}]_{:,i_n}
    =\sum_{(n',m')\in\mathbb{I}_{i_n}}\bm{e}_{n'}^{N_{\tx}} \kron \bm{e}_{m'}^{N_{\rx}}
         = \bm{e}_{n}^{N_{\tx}} \kron \bm{e}_{m}^{N_{\rx}},
\end{align*}
where \ac{rx} sensor index $m$ is fixed for all $n=1,2,\ldots,N_{\tx}$ by Property~\labelcref{i:nrs_rx}. 
Hence, the submatrix of $\bm{\Upsilon}$ formed by columns $\mathbb{J}$ (corresponding to the \ac{nrs} elements $\nrs$) can be written as 
\begin{align*}
     [\bm{\Upsilon}]_{:,\mathbb{J}}&\triangleq 
     \begin{bmatrix}
    [\bm{\Upsilon}]_{:,i_1}&[\bm{\Upsilon}]_{:,i_2}&\ldots&[\bm{\Upsilon}]_{:,i_{N_{\tx}}}
\end{bmatrix}\\
    &=
    \begin{bmatrix}
    \bm{e}_1^{N_{\tx}}\kron\bm{e}_m^{N_{\rx}}&\bm{e}_2^{N_{\tx}}\kron\bm{e}_m^{N_{\rx}}&\ldots &\bm{e}_{N_{\tx}}^{N_{\tx}}\kron\bm{e}_m^{N_{\rx}}
\end{bmatrix}\\
    &=\bm{I}_{N_{\tx}}\kron\bm{e}_m^{N_{\rx}}.
\end{align*}
Similarly, the corresponding columns of $(\bm{S}\kron\bm{I}_{N_{\rx}})\bm{\Upsilon}$ become
\begin{align*}
    (\bm{S}\kron\bm{I}_{N_{\rx}})[\bm{\Upsilon}]_{:,\mathbb{J}}\!=\!
    (\bm{S}\kron\bm{I}_{N_{\rx}})(\bm{I}_{N_{\tx}}\kron\bm{e}_m^{N_{\rx}})\!=\!
    \bm{S}\kron\bm{e}_m^{N_{\rx}}.
\end{align*}
These $N_{\tx}$ columns are linearly dependent when $\bm{S}$ is rank deficient. That is, if $\rank(\bm{S})<N_{\tx}$, then, clearly,
\begin{align*}
    \rank(\bm{S}\kron\bm{e}_m^{N_{\rx}})&=\rank(\bm{S})\rank(\bm{e}_m^{N_{\rx}})
    =\rank(\bm{S})
    <N_{\tx}.
\end{align*}
Hence, $(\bm{S}\kron\bm{I})\bm{\Upsilon}$, which contains the set of linearly dependent columns $(\bm{S}\kron\bm{I})[\bm{\Upsilon}]_{:,\mathbb{J}}$, is also column rank deficient. Thereby \ac{rsc} \eqref{eq:rank_nullity} is violated and the proof is completed.
\end{proof}
A few remarks are in order. Firstly, while the absence of an \ac{nrs} is necessary for satisfying \ac{rsc} (per \cref{thm:unfavorable_set}), it is currently unknown whether it is also sufficient. Secondly, if an \ac{nrs} does not exist, it is still nontrivial to prove that \ac{rsc} is satisfied (and if so, for which choices of rank-deficient  $\bm{S}$). 
Thirdly, finding all array geometries that both have an \ac{nrs} and a contiguous sum co-array is an inherently combinatorial problem, with no know efficient solution. However, to partially bridge this gap, we will now derive one family of such arrays.

\subsection{A family of arrays with \acl{nrs}}
Define the so-called \ac{gma} as
    \begin{align}
     \mathbb{D}_{\tx}\!=\!\{0, \Delta,\ldots,(N_{\tx}\!-\!1)\Delta\},\quad
    \mathbb{D}_{\rx}\!=\!\{0,1,\ldots,N_{\rx}\!-\!1\}.\label{eq:rna}
    \end{align}
Here, $\Delta\leq N_{\rx}$ is a user-defined positive integer. Similar array configurations to \eqref{eq:rna} have featured in other contexts \cite{bliss2007mimo,chen2008mimoradarspace}. However, their ability to operate at reduced \ac{wr} has not been investigated before. 
The following \lcnamecref{thm:gna_subopt} provides conditions under which a GMA permits an NRS and therefore fails to 
satisfy \ac{rsc} at reduced \acl{wr}.
\begin{corollary}[Array geometry with \ac{nrs}]\label{thm:gna_subopt}
    Consider the \ac{gma} in \eqref{eq:rna}. If $N_{\rx}/2<\Delta<N_{\rx}$, then the sum co-array of the \ac{gma} contains a subset 
   $\{\Delta,2\Delta,\ldots,N_{\tx}\Delta\}-1\subset\mathbb{D}_\Sigma$ that forms an \ac{nrs}. Hence, no rank-deficient $\bm{S}$ can satisfy \ac{rsc} \eqref{eq:rank_nullity}.
\end{corollary}
\begin{proof}\let\proof\relax\let\endproof\relax
See supplementary material. 
\end{proof}

An example of a \ac{gma} for $N_{\tx}=7$, $N_{\rx}=6$ and $\Delta = 4$ is
\begin{enumerate}[label=Array \arabic*:,ref=\arabic*,leftmargin=1.7cm,series=arrays]
   \item 
    $\mathbb{D}_{\tx}^{(\aiii)}\!=\!
   \{0,4,8,12,16,20,24\};\!
    \mathbb{D}_{\rx}^{(\aiii)}\!=\!
    \{0,1,\ldots,5\}$,\label{i:array_bad_gma}
 \end{enumerate}
which has a contiguous sum co-array with $N_\Sigma=30$ elements containing \ac{nrs} $\{3,7,11,15,19,23,27\}$. 
While \cref{thm:gna_subopt} specifies a family of array geometries 
that do not permit reduction in \ac{wr}, one can also find arrays 
with favorable redundancy patterns that satisfy \ac{rsc}.
Such an example is 
\begin{enumerate}[label=Array \arabic*:,ref=\arabic*,leftmargin=1.7cm,resume=arrays]
     \item    
    $\mathbb{D}_{\tx}^{(\aiv)}\!=\!
   \{0,6,9,12,15,18,24\};\!
    \mathbb{D}_{\rx}^{(\aiv)}\!=\!
    \{0,1,\ldots,5\}$,\label{i:array_good_2}
 \end{enumerate}
with the same sum co-array and number of physical sensors as \cref{i:array_bad_gma}. It can be verified that \cref{i:array_good_2}, is able to satisfy redundancy subspace condition \eqref{eq:rank_nullity} at reduced \ac{wr}, 
unlike \cref{i:array_bad_gma}, as we will demonstrate numerically in \cref{sec:numerical}. For example, by transmitting 
$N_{\tx}-2$ independent waveforms from Tx sensor subset $\mathbb{D}_{\tx}^{\aiv}\setminus\{9,15\}$,
one can satisfy \ac{rsc}. It can also be verified that the sum coarray of \cref{i:array_good_2} does not contain an \ac{nrs}. Similar conclusions hold for 
 the following two arrays (with identical contiguous co-arrays): 
 \begin{enumerate*}[label=Array \arabic*:,ref=\arabic*,leftmargin=1.7cm,resume=arrays]
     \item    
    $\mathbb{D}_{\tx}^{(\athree)}\!=\!\{0,1,2\}$; $\mathbb{D}_{\rx}^{(\athree)}\!=\!\{0,1,2,5\}$, and \label{i:array_bad_3}
    \item $\mathbb{D}_{\tx}^{(\afour)}\!=\!\mathbb{D}_{\tx}^{(\athree)}$, $\mathbb{D}_{\rx}^{(\afour)}\!=\!\{0,1,3,5\}$.\label{i:array_good_4}
 \end{enumerate*}
Our prior work \cite{rajamaki2023importance} showed that when $\bm{S}$ is rank-deficient, 
$(\bm{S}\kron\bm{I})\bm{\Upsilon}$ has linearly dependent columns in case of \cref{i:array_bad_3} (violating \ac{rsc}), whereas  \cref{i:array_good_4} can satisfy \ac{rsc}. 
This can be clearly understood via  \cref{thm:unfavorable_set} and the \ac{nrs} structure identified herein (unavailable in \cite{rajamaki2023importance}): the sum co-array of \cref{i:array_bad_3} contains an \ac{nrs} (namely, $\{5,6,7\}$), whereas that of \cref{i:array_good_4} does not. 
While this suggests that absence of \ac{nrs} might be both necessary and sufficient for satisfying \ac{rsc}, establishing sufficiency is still an open question and left for future work.

\section{Numerical results}\label{sec:numerical}

Next, we numerically validate the theory derived in \cref{sec:sumcoarray}. 
For \ac{doa} estimation, we recover the co-array (\ac{ula}) measurement via atomic norm minimization \cite{bhaskar2013atomic} by solving 
\begin{align*}
\underset{\bm{z},\bm{u}\in\mathbb{C}^{N_\Sigma}; t\geq 0}{\text{minimize}}\ \|\bm{y}-\bm{W}\bm{z}\|_2^2 +(t+u_1)\tau\
	\text{s.t.}\ 
    \begin{bmatrix}
        \mathcal{T}(\bm{u})&\bm{z}\\
        \bm{z}^{\HT}&t
    \end{bmatrix}\succeq 0,
\end{align*}
and then apply root-MUSIC \cite{barabell1983improving} to Hermitian Toeplitz matrix $\mathcal{T}(\bm{u})\in\mathbb{C}^{N_\Sigma\times N_\Sigma}$ (assuming $K$ to be known)\footnote{Regularization parameter $\tau$ is 
set to the standard deviation of the noise $\tau\!=\!\sigma$ (assumed known). 
When $\sigma\!=\!0$, we constrain $\bm{W}\bm{z}\!=\!\bm{y}$ and set $\tau\!=\!1$.}. 
Noise is 
circularly symmetric Gaussian, $\bm{n}\sim\mathcal{CN}(0,\sigma^2\bm{I})$, and
scattering coefficients are i.i.d. on the complex unit circle, $|x_k|=1/\sqrt{K},\forall k$. \Ac{snr} is defined as $ \mathbb{E}\|\bm{x}\|_2^2/\mathbb{E}\|\bm{\bm{n}}\|_2^2=1/\sigma^2$, and the transmit power is $\|\bm{S}\|_{\F}^2=1$. The ground truth \acp{doa} are $\omega_k=\sin\theta_k = -1+\frac{2k-1}{K}+\varepsilon_k$, where $\varepsilon_k$ is i.i.d. and uniformly distributed in $[-\tfrac{1}{10K},\tfrac{1}{10K}],\forall k$. 

\cref{fig:sim_K} shows the (empirical) \ac{mse} of \cref{i:array_bad_gma,i:array_good_2} for a varying $K$ when using low-rank waveforms ($\rank(\bm{S})=N_{\tx}-2=5$). Results are averaged over $10^3$ Monte Carlo trials (varying $\bm{x}$ and $\bm{\theta}$, no noise) and $10$ random constant-modulus (low-rank) waveforms with i.i.d. phases drawn uniformly from $[0,2\pi]$. 
Clearly, \cref{i:array_bad_gma} is unable to identify as many targets as \cref{i:array_good_2}, as corroborated by \cref{fig:realization}, which shows a sample realization for $K=N_\Sigma/2\!=\!15$. This is consistent with \cref{thm:unfavorable_set}: an \ac{nrs} hampers identifiability at reduced \ac{wr}, even for a contiguous sum co-array.

\begin{figure}
\newcommand{\noisevar}{0e+00}
\newcommand{\nummc}{1000}

\newcommand{\arrayi}{GMA}
\newcommand{\arrayii}{alt}
\newcommand{\Ntnum}{7}
\newcommand{\Nrnum}{6}
\newcommand{\NSigmanum}{30}
\newcommand{\Nsnum}{5}
\newcommand{\Knum}{15}
\newcommand{\sepnum}{1e-01}

    	\centering
        \subfloat[Average performance]{\label{fig:sim_K}
    	\begin{tikzpicture} 
    		\begin{axis}[width=4.5 cm,height=2.7 cm,ylabel={MSE},xlabel= {\# of targets, $K$},
            xmin=1,xmax=\Knum,
            xtick={1,3,...,\Knum},
     ymode=log,
     ytick = {1e-8,1},
     ylabel style = {yshift=-.5cm},
     xlabel style = {yshift=.1cm},
    legend style = {at={(0.5,1.03)},anchor=south,draw=none,fill=none},legend columns=1,legend style={/tikz/every even column/.append style={column sep=0.2cm},font=\scriptsize},
      ]
        


			\addplot[red,thick,draw,mark=o,each nth point={2}]
            table[x=K,y=MSE,y expr=\thisrow{MSE}*\Knum]
            {Data/KvsMSE_mean_GMA_rand_anmrmusic_NSigma_30_Nt_7_Nr_6_Ns_5_eta_\noisevar_K_\Knum_sep_1e-01_mc_\nummc.dat};
            \addlegendentry{\cref{i:array_bad_gma} (with \ac{nrs})}



            

            \addplot[black,thick,draw,mark=+,mark options={solid}] table[x=K,y=MSE,y expr=\thisrow{MSE}*\Knum,each nth point={2}]
            {Data/KvsMSE_mean_alt_rand_anmrmusic_NSigma_30_Nt_7_Nr_6_Ns_5_eta_\noisevar_K_\Knum_sep_1e-01_mc_\nummc.dat};%
            \addlegendentry{\cref{i:array_good_2} (no \ac{nrs})}


    		\end{axis}%
    	\end{tikzpicture}
        }\hfil
        \subfloat[Sample realization ($K=15$)]{
        \begin{tikzpicture}%
        \begin{groupplot}[
        group style={
            group size=1 by 2,
            x descriptions at=edge bottom,
            vertical sep=0.1cm,
        },
        footnotesize,
        width=4.5cm,
        height=2.4cm,
        xlabel={Angle, $\theta$},
        ymin=0,ymax=.55,xmin=-pi/2,xmax=pi/2,xlabel={Angle, $\theta$},xtick={-pi/2,0,pi/2},xticklabels={$-\frac{\pi}{2}$,,$\frac{\pi}{2}$},xlabel shift = {-15 pt},
        ymode=linear,
       title style={yshift=-.4cm},
]
    \nextgroupplot[,legend to name=grouplegend
            ,mark=none,legend style = {draw=none,fill=none,/tikz/every even column/.append style={font=\scriptsize}},legend columns=2,title={\scriptsize \cref{i:array_bad_gma} (with \ac{nrs})}]
        \addplot+[ycomb,gray, mark = square,mark options={scale=1},  thick] table[x=theta,y =sig]{Data/gt_K_15.dat};     
        \addlegendentry{Ground truth}
       \addplot+[ycomb,blue,mark=*,mark options={blue,scale=.6,solid}, very thick,dashed] table[x=theta,y=sig]{Data/estimate_GMA_rand_anmrmusic_NSigma_30_Nt_7_Nr_6_Ns_5_K_15_sep_1e-01_eta_0e+00.dat};
       \addlegendentry{Estimate}
    \nextgroupplot[%
            title={\scriptsize \cref{i:array_good_2} (no \ac{nrs})}, 
            ]
        \addplot+[ycomb,gray, mark = square,mark options={scale=1}, thick] table[x=theta,y =sig]{Data/gt_K_15.dat};     
       \addplot+[ycomb,blue,mark=*,mark options={blue,scale=.6,solid}, very thick,dashed] table[x=theta,y=sig]{Data/estimate_alt_rand_anmrmusic_NSigma_30_Nt_7_Nr_6_Ns_5_K_15_sep_1e-01_eta_0e+00.dat};
    \end{groupplot}
    \node at (group c1r1.north) [anchor=north, yshift=.7cm, xshift=0cm] {\ref*{grouplegend}};
    \end{tikzpicture}\label{fig:realization}
        }
     \caption{
     \ac{doa} estimation using low-rank waveforms (in absence of noise).
     Identifiability is hampered at reduced \acf{wr} when the sum co-array contains a \acf{nrs}. 
     \cref{i:array_bad_gma,i:array_good_2} have identical sum co-arrays and the same number of (physical/virtual) sensors.
     }
    \end{figure}

\cref{fig:sim_snr} shows the \ac{mse} of the two arrays as a function of \ac{snr} for both low-rank and full-rank (orthogonal) omnidirectional waveforms, where 
the $\bm{S}$ satisfies $\bm{S}^{\HT}\bm{S}\!=\!\tfrac{1}{7} \bm{I}_7$ and $\bm{S}^{\HT}\bm{S}\!=\!\tfrac{1}{5}\diag([1, 1,0,1,0,1,1])$, respectively. 
The 
number of targets is $K=15$. 
In the low \ac{wr} case, a gap can be seen between \cref{i:array_bad_gma,i:array_good_2}. This gap grows with \ac{snr}, providing further numerical evidence of the importance of avoiding \acp{nrs} 
when operating at reduced \ac{wr}. 
Interestingly, \cref{i:array_good_2} at reduced \ac{wr} achieves an \ac{mse} comparable to \cref{i:array_bad_gma,i:array_good_2} at full \ac{wr}. This suggests that even with noise, judicious array and waveform design can reduce \ac{wr} without performance loss.


\begin{figure}
    \newcommand{\yscalenum}{15}
    \centering
    \begin{tikzpicture} 
    		\begin{axis}[width=8.5 cm,height= 4.5 cm,ylabel={MSE},xlabel= {SNR (dB)},
    xmin=0,xmax=30,
     ymode=log,
    ytick={1,1e-1,1e-2,1e-3,1e-4,1e-5},
    xtick={0,5,10,15,20,25,30},
    xticklabels={$0$,$5$,$10$,,$20$,$25$,$30$},
    xlabel style = {yshift=.35cm},
    legend style = {at={(0.5,1.03)},anchor=south,draw=none,fill=none},legend columns=3,legend style={/tikz/every even column/.append style={column sep=0.1cm,font=\scriptsize}},
      ]  
         \addlegendimage{empty legend}
        \addlegendentry{\cref{i:array_bad_gma} (with \ac{nrs}):}
            \addplot[red,thick,draw,mark=o,mark options={solid}]
            table[x=SNR,y=MSE,y expr=\thisrowno{1}*\yscalenum]
            {Data/SNRvsMSE_GMA_temp_anmrmusic_NSigma_30_Nt_7_Nr_6_Ns_5_eta_1e-03_K_15_sep_1e-01_equirand_mc_1000.dat};
            \addlegendentry{Low rank $\bm{S}$}
            
			\addplot[red,dashed,very thick,draw,mark=o,mark options={solid}]
            table[x=SNR,y=MSE,y expr=\thisrowno{1}*\yscalenum]
            {Data/SNRvsMSE_GMA_uni_anmrmusic_NSigma_30_Nt_7_Nr_6_Ns_7_eta_1e-03_K_15_sep_1e-01_equirand_mc_1000.dat};
            
           \addlegendentry{Full rank $\bm{S}$}
            
           \addlegendimage{empty legend}
           \addlegendentry{\cref{i:array_good_2} (no \ac{nrs}):}

            \addplot[black,thick,draw,mark=+,mark options={solid}]
            table[x=SNR,y=MSE,y expr=\thisrowno{1}*\yscalenum]
            {Data/SNRvsMSE_alt_maxminsensel_anmrmusic_NSigma_30_Nt_7_Nr_6_Ns_5_eta_1e-03_K_15_sep_1e-01_equirand_mc_1000.dat};
            \addlegendentry{Low rank $\bm{S}$}

            \addplot[black,dashed,very thick,draw,mark=+,mark options={solid}]
            table[x=SNR,y=MSE,y expr=\thisrowno{1}*\yscalenum]
            {Data/SNRvsMSE_alt_uni_anmrmusic_NSigma_30_Nt_7_Nr_6_Ns_7_eta_1e-03_K_15_sep_1e-01_equirand_mc_1000.dat};
            
            \addlegendentry{Full rank $\bm{S}$}
            \end{axis}%
    	\end{tikzpicture}
    \caption{\ac{mse} \emph{vs.} \ac{snr} using low-rank 
    and full-rank waveforms. 
    Sensing performance at reduced \ac{wr} deteriorates if the sum co-array contains an \ac{nrs}. However, judicious redundancy pattern design (avoiding \acp{nrs}) can 
    enable low \ac{wr} designs to match the performance of conventional full \ac{wr} designs.
    }
    \label{fig:sim_snr}
\end{figure}

\section{Conclusion}

This paper examined how array redundancy and reduced \acf{wr} impact identifiability in active sensing. 
Reduced WR can lower hardware costs and transmission time, freeing up spatio-temporal resources for \ac{isac} or enabling beamforming without loss of identifiablity. 
We showed that identifiability at low \ac{wr} critically depends on the redundancy \emph{pattern} linking physical and virtual sensors. 
Hence, arrays with identical co-arrays can exhibit fundamentally different identifiability properties under reduced \ac{wr}. We derived a novel necessary condition for maximizing identifiability at low \ac{wr}, revealing a structure in the sum co-array called \acs{nrs} 
limits the performance of certain redundant array geometries. 
An immediate implication is that array design for reduced \ac{wr} should avoid \acp{nrs}. 
Our results provide new insights for resource-efficient \ac{mimo} system design, motivating future research on redundancy-aware array and waveform design.

\bibliographystyle{IEEEtran}
\bibliography{IEEEabrv,references}

\newpage

\section*{Supplementary material: Proof of 
	\cref{thm:gna_subopt}}\label{sec:proof_nonachievability}
We show that if $1<\tfrac{N_{\rx}}{\Delta}<2$, then set $\mathbb{S}=\{\Delta-1,2\Delta-1,\ldots,N_{\tx}\Delta-1\}$ is an \ac{nrs} of the \ac{gma} \eqref{eq:rna}. 
First, note that the sum co-array $\mathbb{D}_\Sigma=\mathbb{D}_{\tx}+\mathbb{D}_{\rx}$ is contiguous by \eqref{eq:rna} and the fact that $\Delta\leq N_{\rx}$. Furthermore, $\mathbb{S}$ is a subset of $\mathbb{D}_\Sigma$, i.e., 
$\mathbb{S}\subseteq\mathbb{D}_\Sigma=\{0,1,\ldots,N_{\rx}+(N_{\tx}-1)\Delta\}$, since $\Delta\geq 1$ and $N_{\rx}-\Delta\geq 0$. 
Moreover, $|\mathbb{S}|=N_{\tx}$ by construction. Therefore, $\mathbb{S}$ trivially satisfies Property~\labelcref{i:nrs_cardinality} of \cref{def:nrs}. To show that $\mathbb{S}$ also satisfies Properties~\labelcref{i:nrs_redundancy,i:nrs_rx}, note that for a GMA, $d_{\tx}[n]=(n-1)\Delta, d_{\rx}[m]=m-1$ and $d_\Sigma[i]=i-1$. Recalling \eqref{eq:set_redundancy}, the set of \ac{tx}-\ac{rx} sensor indices contributing to the $p\Delta$th virtual sensor, where $1\leq p \leq N_{\tx}$, can therefore be written as
\begin{align*} 
	\begin{aligned}
		\mathbb{I}_{p\Delta}\!=\!\{ (n,m)\!\in\![1,N_{\tx}]\!\times\![1,N_{\rx}] \text{ s.t. } (n\!-\!1)\Delta\!+\!m\!-\!1\!=\!p\Delta\!-\!1\}.
	\end{aligned}
\end{align*}
We now claim that $\mathbb{I}_{p\Delta}$ is a singleton with $m=\Delta,\forall p$, i.e.,
\begin{align} 
	\mathbb{I}_{p\Delta}=\{(p,\Delta) \}.\label{eq:Singleton} 
\end{align}
This claim can be established using the condition $1<\frac{N_{\rx}}{\Delta}<2$. Firstly, it is trivial to see that any $(n,m) \in \mathbb{I}_{p\Delta}$ must satisfy $n\leq p$. We now show via contradiction that $\nexists (n,m) \in \mathbb{I}_{p\Delta}$ with $n\leq p-1.$ Suppose that for some $n\leq p-1$, $(n,m) \in \mathbb{I}_{p\Delta}$. In that case, we have $(n-1)\Delta+m-1=p\Delta -1 \implies m = (p-n+1)\Delta$. Since $m \leq N_{\rx}$, this implies that $\Delta= \frac{m}{p-n+1}\leq \frac{N_r}{p-n+1}\leq \frac{N_{\rx}}{2}.$ This contradicts the assumption that $\frac{N_{\rx}}{\Delta}<2$. Hence $\mathbb{I}_{p\Delta}$ is of the form \eqref{eq:Singleton}, implying that $\mathbb{S}$ also satisfies Properties~\labelcref{i:nrs_redundancy,i:nrs_rx} of \cref{def:nrs}, and is thus an \ac{nrs}.\hfill$\blacksquare$

\end{document}